\documentclass{elsart}
\usepackage{epsfig}
\usepackage{graphics}
\usepackage{amsmath}
\usepackage{amssymb}
\usepackage{graphicx}

\begin{document}
\begin{frontmatter}

\title{Dressed Spin of Polarized $^3$He in a Cell}

\author[uiuc]{P.~H.~Chu\corauthref{*}}\corauth[*]{Corresponding authors:}\ead{pchu@illinois.edu}
\author[uiuc]{A.~M.~Esler},
\author[uiuc]{J.~C.~Peng\corauthref{*}}\ead{jcpeng@illinois.edu},
\author[uiuc]{D.~H.~Beck},
\author[uiuc]{D.~E.~Chandler},
\author[uiuc]{S.~Clayton},
\author[scu]{B.~-Z.~Hu},
\author[cuhk]{S.~Y.~Ngan},
\author[cuhk]{C.~H.~Sham},
\author[cuhk]{L.~H.~So},
\author[uiuc]{S.~Williamson},
\author[uiuc]{J.~Yoder}
\address[uiuc]{Department of Physics, University of Illinois at Urbana-Champaign, Urbana, Illinois 61801, USA}
\address[cuhk]{Department of Physics, The Chinese University of Hong Kong, Hong Kong, China}
\address[scu]{Department of Physics, Soochow University, Taipei, Taiwan}
\date{\today}

\begin{abstract}
We report a measurement of the modification of the effective precession frequency of polarized ${}^{3}$He atoms in response to a dressing field in a room temperature cell. The $^3$He atoms were polarized using the metastability spin-exchange method. An oscillating dressing field is then applied perpendicular to the constant magnetic field. Modification of the $^3$He effective precession frequency was observed over a broad range of the amplitude and frequency of the dressing field. The observed effects are compared with calculations based on quantum optics formalism.
\end{abstract}


%\maketitle
\end{frontmatter}
 A non-zero neutron electric dipole moment(EDM) is direct evidence for 
violations of both parity ($P$) and time-reversal ($T$) 
symmetries~\cite{Ramsey:1950, Landau:1957}.
Assuming $CPT$ invariance, $T$ violation also implies $CP$-violation~\cite{Khriplovich:1997}. 
Observation of a non-zero neutron EDM would provide qualitatively new 
information on the origin of $CP$-violation, since no $CP$ violation has 
ever been found for a baryon or a hadron containing light quarks 
only, like a neutron. 

The most sensitive neutron EDM measurement was carried out at the ILL 
(Institut Laue Langevin) using bottled ultracold neutrons (UCNs) and an 
upper limit of $\left|d_n\right|<2.9\times 10^{-26}$ {\it e} cm ($90\%$ 
 C.L.) was obtained~\cite{Baker:2006ts}. A non-zero neutron EDM will lead to Stark splitting in an electric field. 
In the presence of parallel (antiparallel) magnetic ($B_0$) and electric 
($E$) fields, the Larmor precession frequency ($\omega$) is given by
\begin{align}
\hbar \omega = 2(\mu_B B_0\pm d_n E)
\end{align}
where $\mu_B$ and $d_n$ are the neutron magnetic and electric dipole 
moments, respectively. A shift in the precession frequency which 
correlates with the direction and magnitude of the electric field would be 
a signal for the neutron EDM. Two new neutron EDM 
experiments~\cite{edm:2002, cryoedm:2007} have been proposed using 
a novel technique~\cite{Golub:1994cg} of producing and storing UCNs in 
superfluid $^4$He. 

A major challenge of all EDM experiments is to minimize systematic effects arising from the variation of the magnetic field. The technique of dressed spin has been proposed by Golub and Lamoreaux~\cite{Golub:1994cg} to reduce these systematic effects. A small 
concentration ($X \sim 10^{-10}$) of 
polarized $^3$He atoms would be added to the superfluid helium to act as 
both a spin analyser and a comagnetometer. The relative spin orientation 
between the UCNs and $^3$He determines the rate of the absorption reaction
\begin{equation}
  n+{}^3He\rightarrow p+{}^3H+764~{\rm keV},
\end{equation}
which has a much larger cross section when the 
n$-{}^3$He have a total spin 
$J=0$ compared to $J=1$~\cite{Passel:1966}. In a magnetic field 
$B_0$, the UCN and ${}^3$He spins will precess at their respective Larmor 
frequencies: $\omega_n = \gamma_n B_0$ and $\omega_3 = \gamma_3 B_0$, 
where $\gamma_i$ is the gyromagnetic ratio of each species. The relative 
angle between the UCN and ${}^3$He spin will develop over time ($\gamma_3 
\approx 1.1 \gamma_n$), and the rate of the 
absorption reaction is modulated at the difference of the two spin 
precession frequencies:
\begin{align}
\omega_{rel} = (\gamma_3-\gamma_n)B_0 \approx 0.1\gamma_n B_0
\label{eq:2}
\end{align}
In the presence of a static electric field $E$ parallel to $B_0$, Eq.~\ref{eq:2} gains an additional term proportional to the neutron EDM
\begin{align}
\omega_{rel} = (\gamma_3-\gamma_n)B_0 + 2 d_n E/\hbar.
\label{eq:3}
\end{align}

For typical values of $B_0=10$ mG and $E=30$ KV/cm, Eq.~\ref{eq:3} shows 
that the first term is $\sim 8$ orders of magnitude greater than the 
second term for a $d_n$ of $10^{-27}$e cm. This shows the importance of accurately monitoring the value of $B_0$. An alternative approach is to 
adopt the dressed-spin technique~\cite{Golub:1994cg}. In the presence of an oscillating off-resonance 
magnetic field, $B_d\cos\omega_d t$, perpendicular to $B_0$, the UCN and 
${}^3$He magnetic moments can be effectively modified, and in fact equalized, by the 
dressing field set at the so-called `critical-dressing' 
condition. In the high dressing-field frequency limit ($\omega_d \gg \gamma_i B_0$), it was 
shown~\cite{Cohen-Tannoudji:1965} that $\gamma_i$ becomes 
$\gamma_i^{\prime}$ given by
\begin{align}
\gamma_i^{\prime} &= \gamma_i J_0(x_i), ~~~x_i \equiv \gamma_i B_d/\omega_d
\end{align}
where $J_0$ is zeroth-order Bessel function of the first kind. The 
critical dressing occurs at $\gamma_3^\prime = \gamma_n^\prime$ (or 
$x_3=(\frac{\gamma_3}{\gamma_n})x_n=1.323$), where 
the modified gyromagnetic ratios for neutron and $^3$He are equal. 
Eq.~\ref{eq:3} shows that the EDM signal is then independent of $B_0$ and 
not sensitive to the fluctuation and drift of $B_0$.

Modification of the neutron effective magnetic moment using an oscillating 
magnetic field has been studied by Muskat \textit{et 
al.}~\cite{Muskat:1987} for a polarized neutron beam. More recently, modification of $^3$He effective magnetic moment was studied by Esler \textit{et al.}~\cite{Esler:2007dt} using a polarized $^3$He atomic beam. Both experiments utilized polarized beam, and no measurements have been reported yet for polarized neutrons or $^3$He in a cell. As the proposed neutron EDM experiment will 
utilize polarized $^3$He in a superfluid helium cell, it is of interest to extend the previous study~\cite{Esler:2007dt} to polarized $^3$He stored in a cell. 

In this paper, we report a measurement of response of polarized
$^3$He atoms in a room temperature cell to the application of a dressing 
field. Modification of the $^3$He effective 
precession frequency was observed over a broad range of the 
amplitude and frequency of the 
dressing field. The observed shift in the effective precession frequency 
are found to be in good agreement with theoretical calculations based on 
quantum optics. 

\begin{figure}[b]
\centering
        \includegraphics[height=0.6\textheight,angle=-90]{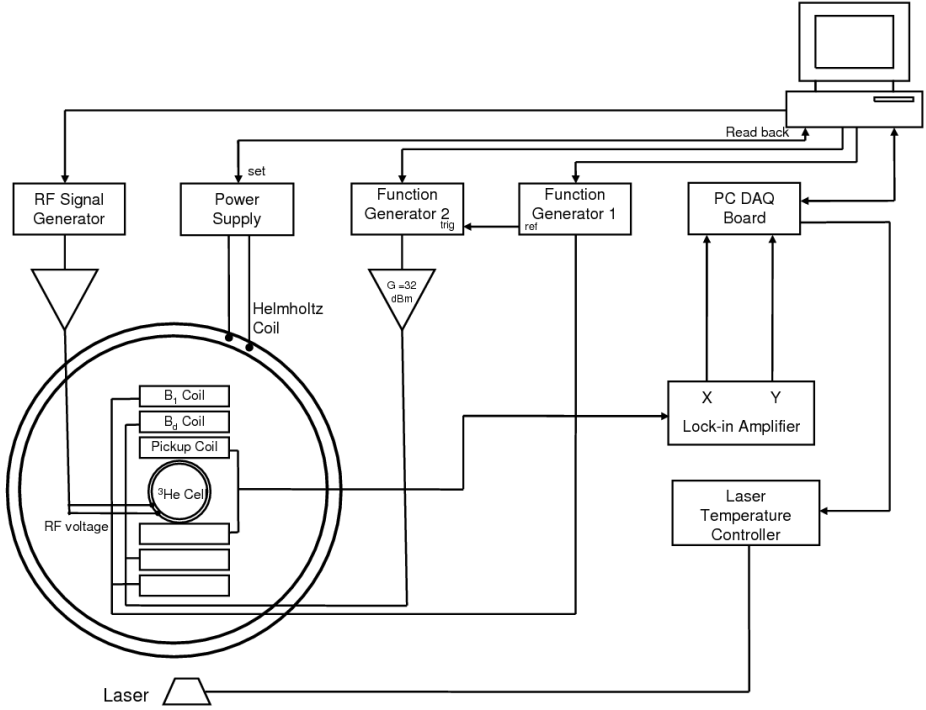}
        \caption{Schematic of the apparatus. See text for description of various components.}
        \label{fig:schematic}
\end{figure}

An overall schematic of the apparatus is shown in 
Fig.~\ref{fig:schematic}. A cylindrical Pyrex cell of 2.5 cm radius and 
5.7 cm length is filled with 1 torr $^3$He gas and located at the center 
of a pair of 50.8 cm radius Helmholtz coils, which provides $B_0$ along 
the $z$-axis. Another pair of Helmholtz coils of 25.4 cm radius cancels 
the vertical component of the Earth field. Four 80-turn
rectangular pickup coils of 5.08 cm $\times$ 6.35 cm are placed in 
the $\hat{x}-\hat{z}$ plane surrounding the cell to measure the $^3$He 
precession signal. Two other pairs of coils, the $B_1$ and the dressing coils, 
provide oscillatory magnetic fields along the $x$-axis. The radius and 
separation of the $B_1$ coils are 11.94 cm and 12.7 cm, respectively. For 
the dressing coils, the radius is 11.94 cm and the separation is 10.8 cm.

The $^3$He gas is polarized using the metastability spin exchange 
method~\cite{Colegrove:1963,Gentile:1993}. An RF field is applied 
to two electrodes outside of the cell to generate a discharge in the 
$^3$He gas. A GaAs diode laser provides circularly polarized light with 
wavelength tuned 
at 1083 nm to pump $^3$He atom from the metastable $2^3 S_1~F=\frac{3}{2}$ 
states to $2^3 P_0~F=\frac{1}{2}$ states 
($C_9$ transitions). A polarization of $\sim 20\%$ for $^3$He 
is obtained~\cite{Williamson:2008}.

After $^3$He is polarized, the laser and RF discharge are turned off, 
followed by the application of a short oscillatory pulse at Larmor 
frequency on the $B_1$ coil 
to rotate the $^3$He spin from $\hat{z}$ to the $\hat{x}-\hat{y}$ plane. 
The dressing 
field $B_d\cos \omega_d t~\hat{x}$ is then applied. Without the dressing 
field, the $^3$He atoms precess at the Larmor frequency $\omega_0 = 
\gamma_3 B_0$. The application of the dressing field modifies the $^3$He 
effective precession 
frequency. 

\begin{figure}[b]
	\centering
        \includegraphics[width=0.85\textwidth,height=0.4\textheight]{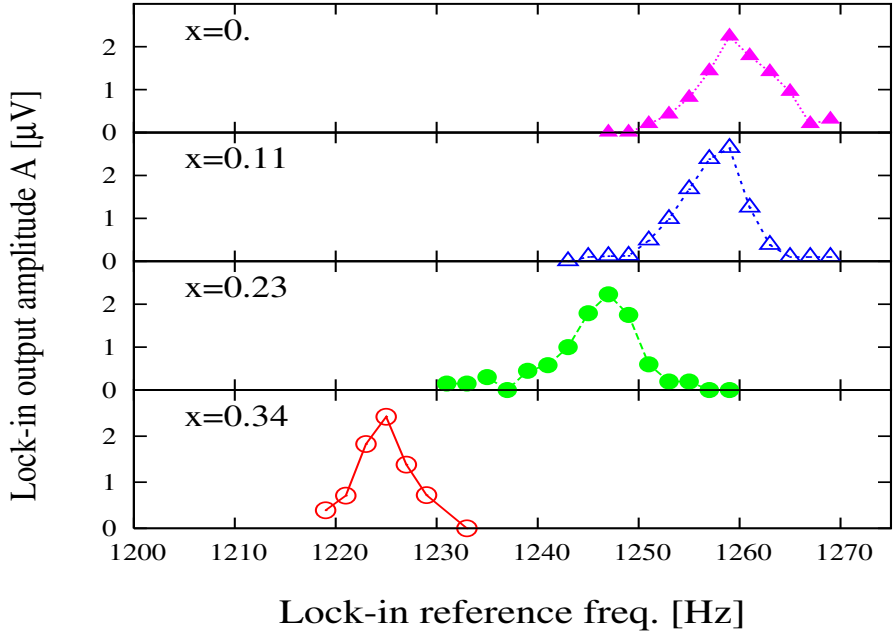}
        \caption{ Examples of amplitudes of lock-in signals vs. reference 
frequency. The resonance frequency shifts with the magnitude of the 
dressing field $B_d$ ($x=\gamma B_d/\omega_d$). For these data $B_0 = 
387.7$ mG, $\omega_0/2\pi = 1257$ Hz, and $\omega_d/2\pi=7152.5$ Hz 
$(y=0.176)$.}
        \label{fig:pickupsignalamplitude}
\end{figure}

The time-varying magnetization caused by the $^3$He precession will induce an EMF in the pickup coils. The signal from the pickup coils is then analyzed by a lock-in amplifier to measure the rms voltage $V(t)$ at the reference frequency of the lock-in. For each setting of the dressing field, the reference frequency is varied to locate the maximum output amplitude, which occurs when the reference frequency coincides with the $^3$He precession frequency. Fig.~\ref{fig:pickupsignalamplitude} shows examples of measurements for several settings of the magnitude of $B_d$ at $B_0=387.7$ mG and $\omega_d/2\pi=7152.5$ Hz. As $B_d$ (or equivalently, $x\equiv \gamma_3 B_d/\omega_d$) increases, the $^3$He precession frequency clearly shifts to lower values. The widths of the peaks in Fig.~\ref{fig:pickupsignalamplitude} are consistent with the 5.3 Hz bandwidth of the lock-in amplifier. The effective precession frequency, $\omega^{eff}$, can be determined from the location of maximum amplitude $A$.

%A typical output from the lock-in amplifier, $V(t)$, is shown in Fig.~\ref{fig:pickupsignal}. The amplitude $A$ is obtained by fitting $V(t)$ with the form $A\exp{(-t/T_2^{*})+C}$.Without the dressing field, the $^3$He spin relaxation time ($T_2^{*}$) was measured to be $\sim 1$ sec, refecting the non-uniformity of the magnetic field. In the presence of the dressing field, $T_2^*$ was found to decrease further presumably due to the additional non-uniformity of the dressing field. 

\begin{table}[t]
		\caption{List of parameters for the various measurements.}
		\centering
\begin{tabular}{|c|c|c||c|c|c|}
\hline
$y$ & $\frac{\omega_0}{2\pi}$ [Hz] & $\frac{\omega_d}{2\pi}$[Hz] &$y$ & 
$\frac{\omega_0}{2\pi}$[Hz]&$\frac{\omega_d}{2\pi}$[Hz]\\
\hline
\hline
0.15& 1271& 8473&1.1& 1267& 1151.8\\
\hline
0.3& 1272& 4240 &1.5& 1267& 844.7\\
\hline
0.5& 1267& 2540 &2.5& 1267& 506.8\\
\hline
0.8& 1271& 1588&4.5& 1267& 281.5\\
\hline
0.9& 1271& 1412&7.5& 1270& 169.33\\
\hline
\end{tabular}
\label{tab:data}
\end{table}

\begin{figure}[b]
\centering
        \includegraphics[width=0.8\textwidth]{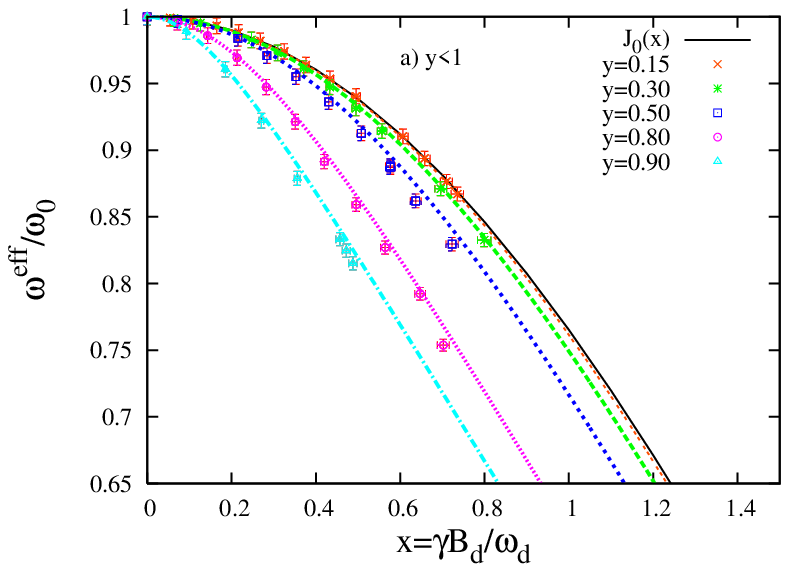}
        \includegraphics[width=0.8\textwidth]{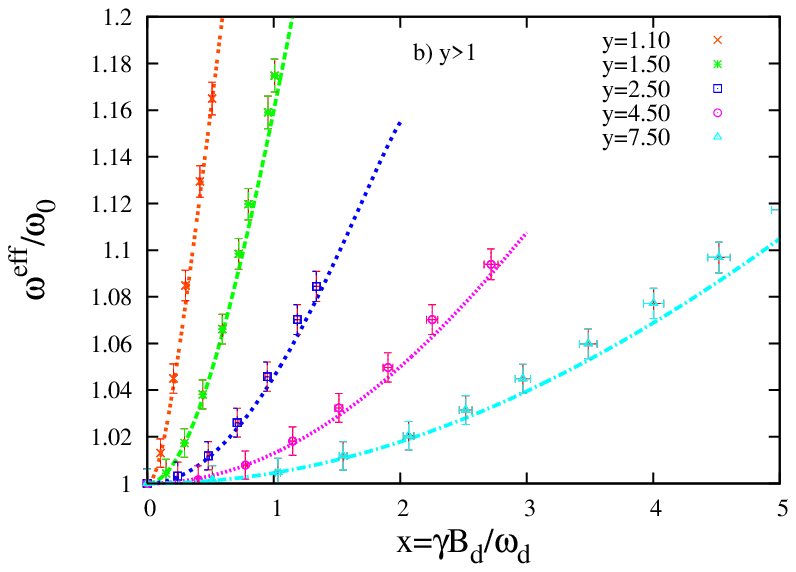}
        \caption{Ratios of measured effective precession frequency 
$\omega^{eff}$ over $\omega_0$ as a function of $x$ for (a)
$y<1$ and (b) $y>1$. The dashed curves are calculations 
described in the text.}
        \label{fig:data}
\end{figure}

The dressed-spin effects are measured for a range of $B_d$ 
and $\omega_d$. 
For convenience, two dimensionless parameters are defined: 
\begin{align}
x&\equiv \frac{\gamma_3 B_d}{\omega_d}, ~~~y\equiv \frac{\gamma_3 
B_0}{\omega_d} = \frac{\omega_0}{\omega_d}.
\end{align}
As will be seen later, the dressed-spin effects are functions fo $x$ and 
$y$ only. Table~\ref{tab:data} lists the values of $B_0$($\omega_0$), 
$\omega_d$ and $y$ for our measurement. Fig.~\ref{fig:data} shows the measured ratios 
$\omega^{eff}/\omega_0$ as a function of $x$ and $y$.
For a given value of $y$, $B_0$ and $\omega_d$ are fixed and $B_d$ is 
varied to determine the $x$-dependence of $\omega^{eff}/\omega_0$. The error bars in 
Fig.~\ref{fig:data} took into account the uncertainties of the $B_d$ calibration, 
the determination of $\omega^{eff}$, and the drift in $B_0$.

Fig.~\ref{fig:data}(a) shows that $\omega^{eff}$ is smaller than $\omega_0$ when the dressing field frequency is higher than the Larmor frequency ($y=\omega_0/\omega_d < 1$). Also shown in Fig.~\ref{fig:data}(a) is the zeroth-order Bessel function $J_0(x)$. The data are consistent with the theoretical prediction~\cite{Cohen-Tannoudji:1965} that $\omega^{eff}/\omega_0 = \gamma^\prime/\gamma = J_0(x)$ as $y \rightarrow 0$. Indeed, the $\omega^{eff}/\omega_0$ data obtained at $y=0.15$ are well described by 
$J_0(x)$. As $y$ increases toward 1, large deviations from $J_0(x)$ are observed for $\omega^{eff}/\omega_0$, as shown in Fig.~\ref{fig:data}(a).

Fig.~\ref{fig:data}(b) shows $\omega^{eff}/\omega_0$ measured at five different values of $y$ when the dressing field frequencies are lower than the Larmor frequency, namely $y = \omega_0/\omega_d > 1$. In contrast to the results observed in Fig.~\ref{fig:data}(a), the dressing frequency $\omega^{eff}$ is now larger than $\omega_0$. In the remainder 
of this paper, we discuss the theoretical calculations for interpreting the observed dressed-spin effects and compare the data with the calculations. 

\begin{figure}[b]
\centering
\includegraphics[width=0.85\textwidth, height=0.43\textheight,angle = 0]{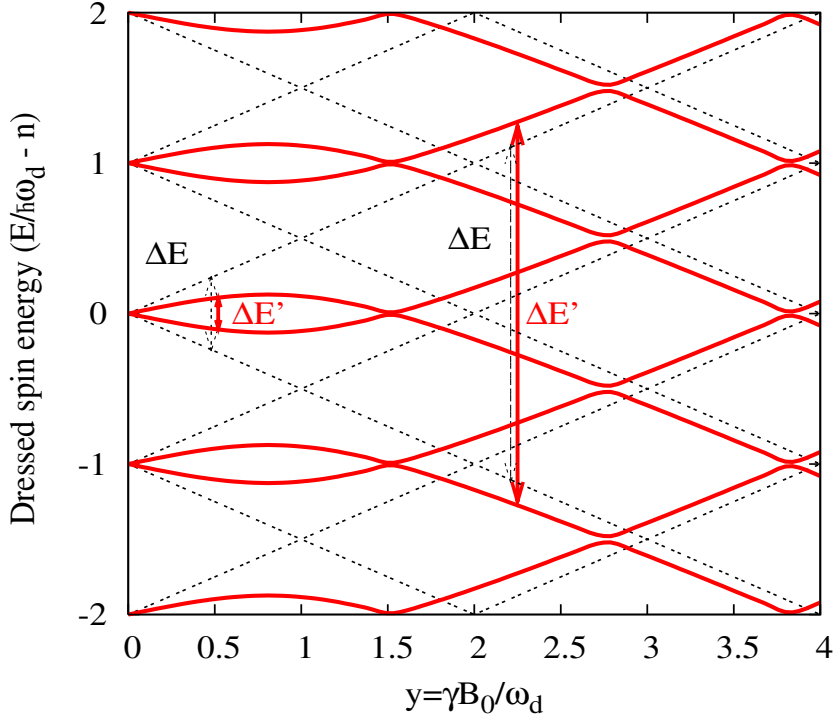}
\caption{ 
Energy diagram of the dressed 
spin system calculated as a function of $y$, for dressing parameter 
$x=1.323$. 
Dashed lines indicate the Zeeman splittings in the undressed system 
($E_0 = \pm \frac{1}{2}\hbar\omega_0$). The solid lines show the 
modified energy spectrum due to the dressing field. The energy scale is 
given in units of the dressing field photon energy $\hbar \omega_d$.}
\label{fig:eigenvalue}
\end{figure}

The dressed spin system was first studied by Cohen-Tannoudji 
{\it et al.}~\cite{Cohen-Tannoudji:1965, Haroche:1970b}. The Hamiltonian 
for a spin-$\frac{1}{2}$ particle with gyromagnetic 
ratio $\gamma$ in a constant magnetic field $B_0\hat{z}$ and an 
oscillatory magnetic field $B_d\cos{\omega_dt}~\hat{x}$ can be written as
\begin{align}
  \frac{\mathcal{H}}{\hbar \omega_d} &= (\frac{\gamma B_0}{\hbar 
\omega_d})\hat{S}_z + 
\hat{a}^{\dagger}\hat{a}+\frac{\lambda}{\hbar\omega_d}\hat{S}_x(\hat{a}+
\hat{a}^\dagger) \notag \\
  & \equiv \frac{y}{2}\hat{\sigma}_z+ \hat{a}^{\dagger}\hat{a} + 
\frac{x}{4\sqrt{\bar{n}}}\hat{\sigma}_x (\hat{a}+\hat{a}^\dagger),
  \label{eq:hamil}	
\end{align}
where $\hat{S}_x$ and $\hat{S}_z$ are the spin operators along 
$\hat{x}$ and $\hat{z}$, respectively. The first term in 
Eq.~\ref{eq:hamil} is 
the Zeeman interaction of the spin with $B_0$, and the second term is the 
energy of the oscillatory dressing field with creation and annihilation 
operators $\hat{a}^\dagger$ and $\hat{a}$. The final term in 
Eq.~\ref{eq:hamil} describes the coupling between the spin of the 
particle and the dressing field with strength 
$\lambda = \gamma B_d/2\sqrt{\bar{n}}$, where $\bar{n}\gg 1$ is the 
average number of photons. This interaction term allows the particle to 
absorb or emit photons and exchange energy and angular momentum with 
the dressing field. Because the dressing field is perpendicular to $B_0$ 
and can be 
decomposed into a superposition of right- and left-handed circularly 
polarized fields, only $\Delta m_z = \pm \hbar$ transitions are allowed.

In the high dressing field frequency limit ($\omega_d \gg 
\gamma B_0$ or $y \ll 1$), 
Eq.~\ref{eq:hamil} can be solved analytically with the result 
$\gamma^{\prime} = \gamma J_0(x)$ ~\cite{Cohen-Tannoudji:1965}. The 
precession frequency becomes
\begin{align}
  \frac{\omega^{eff}}{\omega_0} &= \frac{\gamma^{\prime} B_0}{\gamma 
B_0} = J_0(x),
\end{align}
which only depends on the dressing strength $x=\gamma B_d/\omega_d$. 
The modification of $\gamma$ by a factor of $J_0(x)$ under a high 
frequency dressing field was observed in several 
experiments~\cite{Muskat:1987,Esler:2007dt} including the 
present measurement (see Fig.~\ref{fig:data}). 

To understand the observed deviation of $\gamma^{\prime}/\gamma$ from 
$J_0(x)$ in Fig.~\ref{fig:data} for arbitrary dressing parameters $x$ and 
$y$, we have calculated the energy level diagrams for the dressed-spin 
system by diagonalizing the
Hamiltonian of 
Eq.~\ref{eq:hamil}~\cite{Yabuzaki:1974}. In the 
basis of 
$ ( \cdots, 
\left| n+1, - \right\rangle ,
\left| n,+ \right\rangle ,
\left| n,- \right\rangle ,
\left| n-1,+ \right\rangle, 
\cdots )$, where $n$ signifies the oscillating quanta of the dressing 
field and $+/-$ 
denotes the spin up/down state of $^3$He, the Hamiltonian has the 
following matrix elements: 
\\
%\tiny
\begin{equation}
\left(
\begin{array}{cccccccc}
\cdot & \cdot & \cdot & \cdot & \cdot & \cdot & \cdot & \cdot \\
\cdot & n+1+\frac{y}{2}  & 0 & 0 & \frac{x}{4} & 0 & 0 &  \cdot \\
\cdot & 0 & n+1-\frac{y}{2} & \frac{x}{4} & 0 & 0 & 0 & \cdot \\
\cdot & 0 & \frac{x}{4} & n+\frac{y}{2} & 0 & 0 & \frac{x}{4} &\cdot \\
\cdot & \frac{x}{4} & 0 & 0 & n-\frac{y}{2} & \frac{x}{4}& 0 & \cdot \\
\cdot & 0 & 0 & 0& \frac{x}{4} & n-1+\frac{y}{2} & 0 & \cdot \\
\cdot & 0 & 0 & \frac{x}{4} & 0 & 0 & n-1-\frac{y}{2} & \cdot \\
\cdot & \cdot & \cdot & \cdot & \cdot & \cdot & \cdot & \cdot
\end{array}
\right)
%\left(
%\begin{array}{c}
%\cdot \\
%\left| +, n+1 \right\rangle \\ 
%\left| -, n+1 \right\rangle \\ 
%\left| +, n   \right\rangle \\ 
%\left| -, n   \right\rangle \\ 
%\left| +, n-1 \right\rangle \\ 
%\left| -, n-1 \right\rangle \\ 
%\cdot
%\end{array}
%\right)
\end{equation}
\normalsize

Fig.~\ref{fig:eigenvalue} shows an example of the energy eigenvalue 
diagram as a function of the parameter $y=\gamma_3
B_0/\omega_d$ for a dressing field magnitude corresponding to 
$x=\gamma_3 B_d /\omega_d = 1.323$, which corresponds to the condition 
for critical dressing. A matrix of dimension $46\times 46$ is diagonalized to 
obtain the energy eigenvalues for each $x$ and $y$ parameters. 
The dashed lines are the Zeeman splitting for the undressed system ($x=0$).
 Fig.~\ref{fig:eigenvalue} 
shows how the Zeeman splitting in the undressed system is
modified by the presence of the dressing field.
Without the dressing field, the gyromagnetic ratio is given by $\Delta 
E/\hbar B_0$, which is just a constant ($\Delta E/\hbar B_0 = \gamma$) 
independent of $B_0$. When the dressing field is applied, 
Fig.~\ref{fig:eigenvalue} shows that $\Delta E$ is changed to $\Delta 
E^{\prime}$ and $\gamma$ now becomes $\gamma^\prime =
\Delta E^\prime/\hbar B_0$. It is interesting to note that for $y<1$, 
$\Delta 
E^\prime < \Delta E$ and $\gamma^\prime$ is smaller than $\gamma$. In 
contrast, for $y > 1$, Fig.~\ref{fig:eigenvalue} shows that $\Delta 
E^\prime > \Delta E$ and $\gamma^\prime$ is now greater than $\gamma$. 
The striking feature observed in the data shown in 
Fig.~\ref{fig:data}, namely, $\omega^{eff}/\omega_0 = \gamma^\prime 
/\gamma < 1$ for $y<1$ and $\omega^{eff}/\omega_0 > 1$ for $y>1$, is
 well described by this approach.

The dashed curves in Fig.~\ref{fig:data} are the calculations for 
$\omega^{eff}/\omega_0 = \gamma^\prime/\gamma$ using the quantum mechanical method.
 The good agreement between the data and the calculation shows that the 
observed deviation can be quantitatively described in this quantum 
mechanical approach.

In summary, we have developed the method to measure the modification of 
the effective precession frequency of polarized $^3$He atoms in a room temperature cell over a broad range of 
the dressing field parameters. Our study confirms that in the high-frequency limit, the modified 
gyromagnetic ratio $\gamma^\prime$ obeys the relation $\gamma^\prime = 
\gamma J_0(x)$. Deviation from this relation was observed for other 
settings of the dressing field parameters. In particular, we found that 
when $y>1$, $\gamma^\prime$ is larger than $\gamma$. The observed modification
 of the effective $^3$He precession frequency can be quantitatively described by a quantum approach. 
We plan to extend the measurements to cover the larger $x$ region as well
as in superfluid $^4$He cell, which are relevant to the future neutron EDM experiment.

We gratefully acknowledge valuable discussions with M.~Hayden, 
T.~Gentile, R.~Golub, 
and S.~Lamoreaux. This work was supported in part by the National Science 
Foundation.


%\bibliographystyle{elsarticle-num}

\end{document}